\documentclass[prl,aps,twocolumn,superscriptaddress,showpacs,preprintnumbers,asmath,amssymb]{revtex4}
\usepackage{color}
\usepackage{graphicx}% Include figure files
\usepackage{ulem}
\usepackage{float}
\usepackage{here}
\usepackage{hyperref}
\def\textbf#1{\boldsymbol{#1}}

\begin{document}

\title{Momentum-dependent charge correlations in YBa$_2$Cu$_3$O$_{6+\delta}$
superconductors probed by resonant x-ray scattering: Evidence for three competing phases}

\author{S.~Blanco-Canosa}
\affiliation{Max-Planck-Institut~f\"{u}r~Festk\"{o}rperforschung,
Heisenbergstr.~1, D-70569 Stuttgart, Germany}

\author{A.~Frano}
\affiliation{Max-Planck-Institut~f\"{u}r~Festk\"{o}rperforschung,
Heisenbergstr.~1, D-70569 Stuttgart, Germany}
\affiliation{Helmholtz-Zentrum Berlin f\"{u}r Materialien und Energie,
Albert-Einstein-Strasse 15, D-12489 Berlin, Germany}

\author{T.~Loew}
\affiliation{Max-Planck-Institut~f\"{u}r~Festk\"{o}rperforschung,
Heisenbergstr.~1, D-70569 Stuttgart, Germany}

\author{Y.~Lu}
\affiliation{Max-Planck-Institut~f\"{u}r~Festk\"{o}rperforschung,
Heisenbergstr.~1, D-70569 Stuttgart, Germany}

\author{J.~Porras}
\affiliation{Max-Planck-Institut~f\"{u}r~Festk\"{o}rperforschung,
Heisenbergstr.~1, D-70569 Stuttgart, Germany}

\author{G.~Ghiringhelli}
\affiliation{CNR-SPIN, CNISM and Dipartimento di Fisica,
Politecnico di Milano, piazza Leonardo da Vinci 32, I-20133 Milano, Italy}

\author{M.~Minola}
\affiliation{CNR-SPIN, CNISM and Dipartimento di Fisica,
Politecnico di Milano, piazza Leonardo da Vinci 32, I-20133 Milano, Italy}

\author{C.~Mazzoli}
\affiliation{CNR-SPIN, CNISM and Dipartimento di Fisica,
Politecnico di Milano, piazza Leonardo da Vinci 32, I-20133 Milano, Italy}

\author{L.~Braicovich}
\affiliation{CNR-SPIN, CNISM and Dipartimento di Fisica,
Politecnico di Milano, piazza Leonardo da Vinci 32, I-20133 Milano, Italy}

\author{E.~Schierle}
\affiliation{Helmholtz-Zentrum Berlin f\"{u}r Materialien und Energie,
Albert-Einstein-Strasse 15, D-12489 Berlin, Germany}

\author{E.~Weschke}
\affiliation{Helmholtz-Zentrum Berlin f\"{u}r Materialien und Energie,
Albert-Einstein-Strasse 15, D-12489 Berlin, Germany}

\author{M.~Le Tacon}
\email[Corresponding author: ]{m.letacon@fkf.mpg.de}
\affiliation{Max-Planck-Institut~f\"{u}r~Festk\"{o}rperforschung,
Heisenbergstr.~1, D-70569 Stuttgart, Germany}

\author{B.~Keimer}
\email[Corresponding author: ]{b.keimer@fkf.mpg.de}
\affiliation{Max-Planck-Institut~f\"{u}r~Festk\"{o}rperforschung,
Heisenbergstr.~1, D-70569 Stuttgart, Germany}

\date{\today}

\begin{abstract}
%To gain insight into the influence of disorder on the electronic phase behavior of underdoped high-temperature superconductors,
We have used resonant x-ray scattering to determine the momentum dependent charge correlations in YBa$_2$Cu$_3$O$_{6.55}$ samples with highly ordered chain arrays of oxygen acceptors (ortho-II structure). The results reveal nearly critical, biaxial charge density wave (CDW) correlations at in-plane wave vectors (0.315, 0) and (0, 0.325). The corresponding scattering intensity exhibits a strong uniaxial anisotropy. The CDW amplitude and correlation length are enhanced as superconductivity is weakened by an external magnetic field. Analogous experiments were carried out on a YBa$_2$Cu$_3$O$_{6.6}$ crystal with a dilute concentration of spinless (Zn) impurities, which had earlier been shown to nucleate incommensurate magnetic order. Compared to pristine crystals with the same doping level, the CDW amplitude and correlation length were found to be strongly reduced. These results indicate a three-phase competition between spin-modulated, charge-modulated, and superconducting states in underdoped YBa$_2$Cu$_3$O$_{6+\delta}$.
\end{abstract}


%%%%%%%%%%%%%%%%%%%%%%%%%%%%INTRODUCTION%%%%%%%%%%%%%%%%%%%%%%%%%%%%%%%%%%%%%%%%%%%%%%%%%%%%%%%%
%%%%%%%%%%%%%%%%%%%%%%%%%%%%%%%%%%%%%%%%%%%%%%%%%%%%%%%%%%%%%%%%%%%%%%%%%%%%%%%%%%%%%%%%%%%%%%%%

\maketitle

High-temperature superconductivity in the cuprates arises from doping of charge carriers into Mott-insulators with antiferromagnetically ordered CuO$_2$ planes.~\cite{Lee_RMP06} At sufficiently high density, the carriers screen out the random potential created by the donor or acceptor atoms and generate a uniform metallic state out of which superconductivity arises. In underdoped cuprates, however, the screening is less effective, and the role of materials-specific disorder in the formation of the unusual spin and/or charge textures observed in this regime of the phase diagram has been a subject of long-standing debate.~\cite{Kapitulnik_NJP2009, Hosur_2012,Hinkov_Science2008,Fujita_JPSJ2012,Kivelson_RMP2003, Votja_AdvPhys2009,Fink_PRB2009,Fink_PRB2011,Wilkins_PRB2011,Abbamonte_NatPhys2005,Hucker_PRB2011,Wu_NatureCommunications2012,Wise_NPhys2008,Koshaka_Science2007,Parker_Nature2010}
Recent research on the stoichiometric underdoped compounds YBa$_2$Cu$_3$O$_{6.5}$ and YBa$_2$Cu$_4$O$_8$ has provided new perspectives for the resolution of the influence of disorder on the electronic phase behavior of the underdoped cuprates. In these materials, the oxygen acceptors are arranged in ordered chains rather than placed randomly in the crystal lattice, so that chemical and structural disorder is minimized.~\cite{Liang_PhysicaC2000} The results of recent quantum oscillation experiments~\cite{Bangura_PRL2008,Sebastian_RPP2012,Doiron_Nature2007} in high magnetic fields indicate a reconstruction of the Fermi surface by a long-range electronic superstructure.~\cite{Laliberte_NatCom2011,Sebastian_PRL2012,Leboeuf_PRB2011} This discovery has sparked another intense debate on the nature of the high-field ordering and its relation to the ``pseudogap'' observed in these and other underdoped cuprates~\cite{Timusk_RPP99} above the superconducting transition temperature, $T_c$, in the absence of external fields. The pseudogap, in turn, is intimately related to the superconducting gap, and an explanation of its origin is considered an essential element of any theory of high-temperature superconductivity.

Whereas research on YBa$_2$Cu$_4$O$_8$ has been limited, because only small crystals are available and the doping level cannot be varied in a straightforward manner, YBa$_2$Cu$_3$O$_{6.5}$ is a member of the extensively studied  YBa$_2$Cu$_3$O$_{6+\delta}$ (YBCO$_{6+\delta}$, 123) family, where the concentration of mobile holes in the CuO$_2$ layers can be controlled via the oxygen content $\delta$. It was recently shown that compounds with $\delta \sim 0.5$, which are in the focus of current attention because of their minimal disorder and high-field quantum oscillations, mark the boundary between two fundamentally different regimes of the phase diagram.~\cite{Sebastian_PNAS2010,Baek_PRB2012} For $\delta < 0.5$, uniaxial incommensurate magnetic order forms gradually upon cooling and coexists with superconductivity at low temperatures.~\cite{Hinkov_Science2008,Haug_NJP2010} Unlike the well known ``striped'' state in the (La,Nd,Eu)$_{2-x}$(Ba,Sr)$_x$CuO$_4$ (214) family,~\cite{Votja_AdvPhys2009,Fink_PRB2009,Fink_PRB2011,Wilkins_PRB2011,Abbamonte_NatPhys2005,Hucker_PRB2011} which encompasses both spin and charge degrees of freedom, a corresponding charge modulation has not been found in the low-doping regime of YBCO$_{6+\delta}$. For $\delta > 0.5$, on the other hand, static magnetic order disappears, and the magnetic excitation spectrum determined by neutron scattering exhibits a large spin gap.~\cite{Fong_PRB2000} Recent resonant~\cite{Achkar_PRL2012,Ghiringhelli_Science2012} and nonresonant~\cite{Chang_NaturePhysics2012} x-ray diffraction experiments on samples in this doping regime have revealed nearly critical, biaxial charge density wave (CDW) correlations. High-field nuclear magnetic resonance (NMR) experiments~\cite{Wu_Nature2011} for $\delta \sim 0.5$ have been interpreted in terms of a uniaxial commensurate charge-ordered state akin to the commensurate striped state observed in the 214 family for $x = 1/8$,~\cite{Fink_PRB2009,Fink_PRB2011,Wilkins_PRB2011,Abbamonte_NatPhys2005,Hucker_PRB2011} which is incompatible with the wavevector extracted from the x-ray experiments at higher $\delta$. Since neutron or x-ray scattering evidence for spin or charge modulations has not yet been reported for YBCO$_{6.5}$ samples with well-ordered arrays of alternating full and empty oxygen chains (``ortho-II'' structure), the origin of this discrepancy -- and hence the spin and/or charge modulation pattern that generates the Fermi surface reconstruction~\cite{Sebastian_RPP2012,Laliberte_NatCom2011} -- have remained unresolved. In particular, the influence of random disorder created by oxygen defects in samples with off-stoichiometric composition ($\delta \neq 0.5$) on the electronic phase behavior has yet to be clarified.

In order to address these issues, we have carried out complementary resonant x-ray scattering (RXS) experiments on fully untwinned, stoichiometric, highly ortho-II ordered YBCO$_{6.55}$ single crystals with minimal chemical and structural disorder, and on untwinned YBCO$_{6.6}$ crystals in which a controlled amount of disorder was generated by replacing magnetic copper ions in the CuO$_2$ planes by spinless (Zn) impurities. Single crystals of YBCO$_{6.55}$ ($T_c$= 61 K), YBCO$_{6.6}$ ($T_c$= 61 K), and YBa$_2$(Cu$_{0.98}$Zn$_{0.02}$)$_3$O$_{6.6}$ (YBCO$_{6.6}$:Zn, $T_c$= 32 K) were synthesized by a self-flux method following previous reports.~\cite{Hinkov_Science2008,Haug_NJP2010,Suchaneck_PRL2010} The oxygen content was controlled by annealing in well defined oxygen partial pressure, and the c-axis lattice parameters were used
to determine the hole-doping levels.~\cite{Lindemer_JACS1989, Liang_PRB2006} The ortho-II sample was prepared following the procedure given in Ref.~\onlinecite{Liang_PhysicaC2000}, and the oxygen content was estimated as 6.55$\pm$ 0.01.
All crystals were mechanically detwinned by heating under uniaxial stress. Zero field RXS measurements were performed in the UHV diffractometer at the UE46-PGM1  beam line of the Helmholtz-Zentrum Berlin at BESSY II, with the electric field of the photons perpendicular to the scattering plane. Magnetic field dependent measurements (up to 6 T) were performed in the high-field diffractometer at the same beamline. The field was applied at an angle of $11.5^{\circ}$ to the $c$-axis, that is, nearly perpendicular to the CuO$_2$ planes. The data analysis was performed by subtracting a background measured at temperature $T = 160$ K for each scan, and fitting to Lorentzian functions. The background was seen to be independent of applied magnetic field. The components of the momentum transfer ${\bf Q} = (H, K, L)$ are quoted in reciprocal lattice units (r.l.u.) of the orthorhombic crystal structure. The YBCO$_{6.55}$ crystals show half-order peaks at ${\bf Q} = (0.5, 0, L)$ with correlation lengths $\sim 100$ {\AA}, as determined by hard x-ray diffraction.

\begin{figure}
\includegraphics[width=0.95\linewidth]{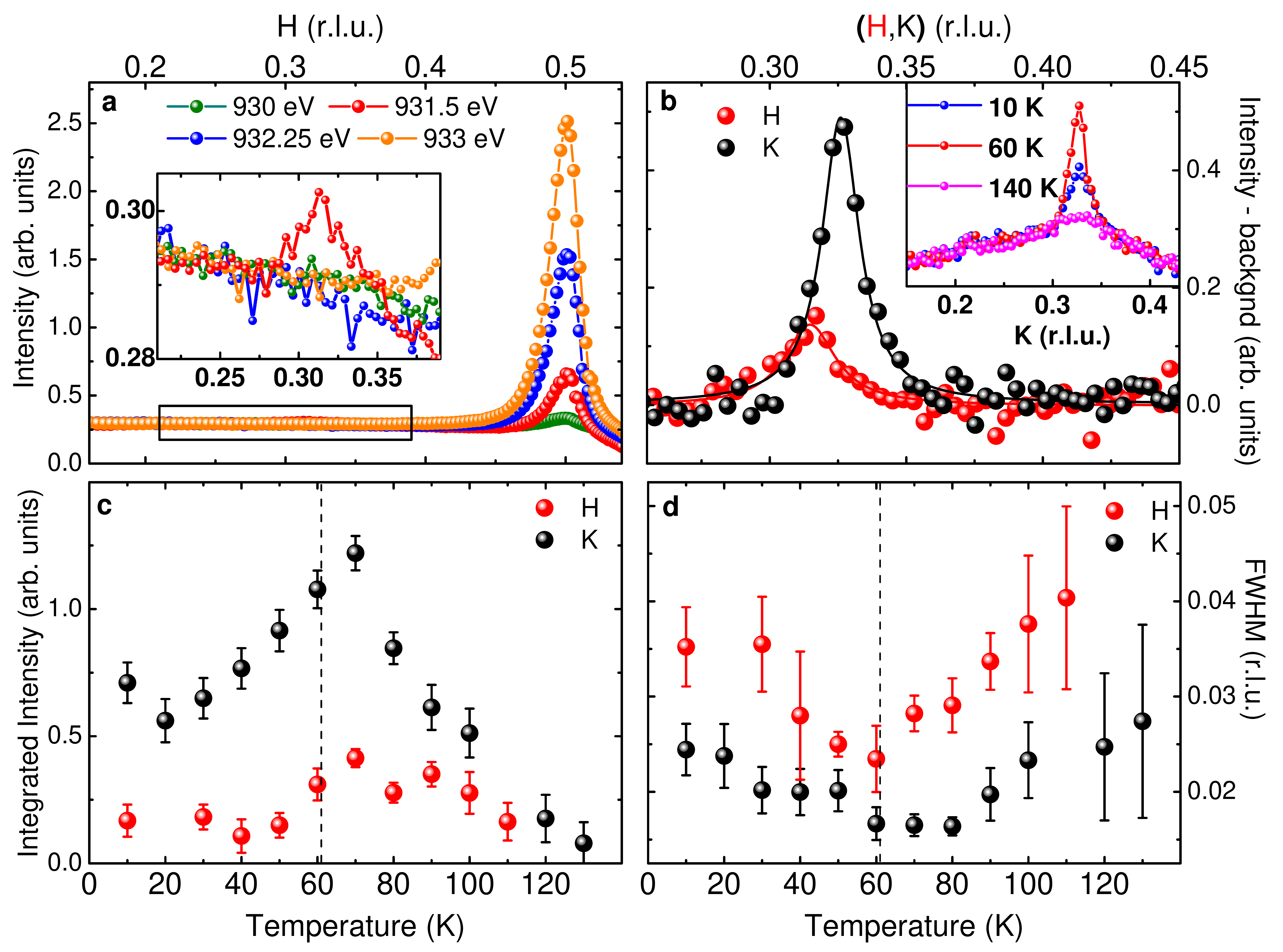}
\caption{(Color online). (a) Scan along the $(H, 0, 0)$ direction for an ortho-II ordered YBCO$_{6.55}$ crystal with $T_c$= 61 K. The intensity of the half-order peak is higher at photon energy 933 eV (resonant conditions for Cu$^+$ in the CuO chains). Inset: Zoom-in around 0.3 r.l.u., showing a weak CDW peak resonating at 931.5 eV. The corresponding $L$-value is $\sim 1.45$, close to the peak maximum, $L = 1.5$, \cite{Chang_NaturePhysics2012} which is outside the Ewald sphere at this photon energy. (b) Anisotropy of the CDW correlations, as revealed by comparing $H$- and $K-$scans. Inset: Temperature dependence (raw data) of the CDW along $(0, K, 0)$. The sloping background is due to the grazing-incidence scattering geometry. Temperature dependence of (c) the integrated intensities and (d) the FWHM of the CDW peaks along $H$ and $K$. Dashed lines indicate $T_c$.
} \label{Fig1}
\end{figure}

Figure~\ref{Fig1}(a) shows the scattered intensity along the $(H, 0, 0)$ direction of reciprocal space (perpendicular to the oxygen chains) for several incident photon energies at $T= T_c$ in YBCO$_{6.55}$. The intense peak seen for $H = 0.5$ at the $L$-absorption edge of the Cu$^{+}$ ions in the oxygen chains (photon energy 933 eV) corresponds to the ortho-II superlattice peak.~\cite{Achkar_PRL2012,Hawthorn_PRB2011} The strong azimuthal dependence of the Bragg intensity confirmed the good orientation of the crystal. The peak is not present above background level along the $(0, K, 0)$ direction, confirming the high detwinning ratio of our sample.
%The presence of a full-empty oxygen pattern in the chains along $b$ direction gives rise to superlattice reflections at (0.5,0,L) rlu, but not at (0,0.5,L), ruling out twinning as responsible for the anisotropy reported here.
The intensity of the half-order peak decreases strongly when tuning the energy to the in-plane Cu$^{2+}$ $L_3$ edge, where we now observe a weak feature at incommensurate in-plane momentum $(q_1, 0)$ with $q_1 = 0.315$ (931.5 eV, inset Fig. 1(a)). Unlike the ortho-II structural peak, rotations of $\pm 3$ degrees around the (0 0 1) direction do not modify the intensity of the charge peak, indicating weaker transverse correlation for this feature.
As in previous RXS reports in the 123 and 214 families,~\cite{Abbamonte_NatPhys2005,Wilkins_PRB2011,Wu_NatureCommunications2012,Achkar_PRL2012,Ghiringhelli_Science2012,Fink_PRB2009,Fink_PRB2011} the energy and polarization dependence of the incommensurate peak (not shown here) confirm that it arises from a charge modulation in the CuO$_2$ planes. \cite{Achkar_PRL2012,Ghiringhelli_Science2012} Its intensity and correlation length peak at $T = T_c$ (see Fig.~\ref{Fig1}(c,d)), indicating that the charge modulation competes with superconductivity, as it does at higher doping levels.~\cite{Achkar_PRL2012,Ghiringhelli_Science2012,Chang_NaturePhysics2012} The low intensity of the feature, which is barely visible above the background, might be responsible for the previously reported null results for samples in the same doping range.~\cite{Ghiringhelli_Science2012,Hawthorn_PRB2011}

As seen in Fig.~\ref{Fig1}(b), a peak is also clearly visible along the $(0, K, 0)$ direction for $K = q_2 = 0.325$. The figure shows that $q_2$ is larger than $q_1$, and that the amplitude of the corresponding peak exceeds the one along the $H$-direction by about a factor of four. The correlation lengths $\xi_{(a,b)} = (a,b)/(\pi*\textrm{FWHM})$ (where $a$ and $b$ are the in-plane lattice parameters, and FWHM is the full with at half maximum of the Lorentzian profile) deduced from these measurement are also different: $\xi_a \sim$ 48 ~\AA, and $\xi_b \sim 82$ ~\AA. The latter value is larger than the correlation lengths reported earlier for both $H$- and $K$-directions in YBCO$_{6.6}$ ($\xi \sim$ 60 \AA, Ref. \onlinecite{Ghiringhelli_Science2012}) and YBCO$_{6.75}$~($\xi \sim$ 40~\AA, Ref. \onlinecite{Achkar_PRL2012}), probably as a consequence of the larger average oxygen chain length in the well-ordered ortho-II structure.

\begin{figure}
\includegraphics[width=0.95\linewidth]{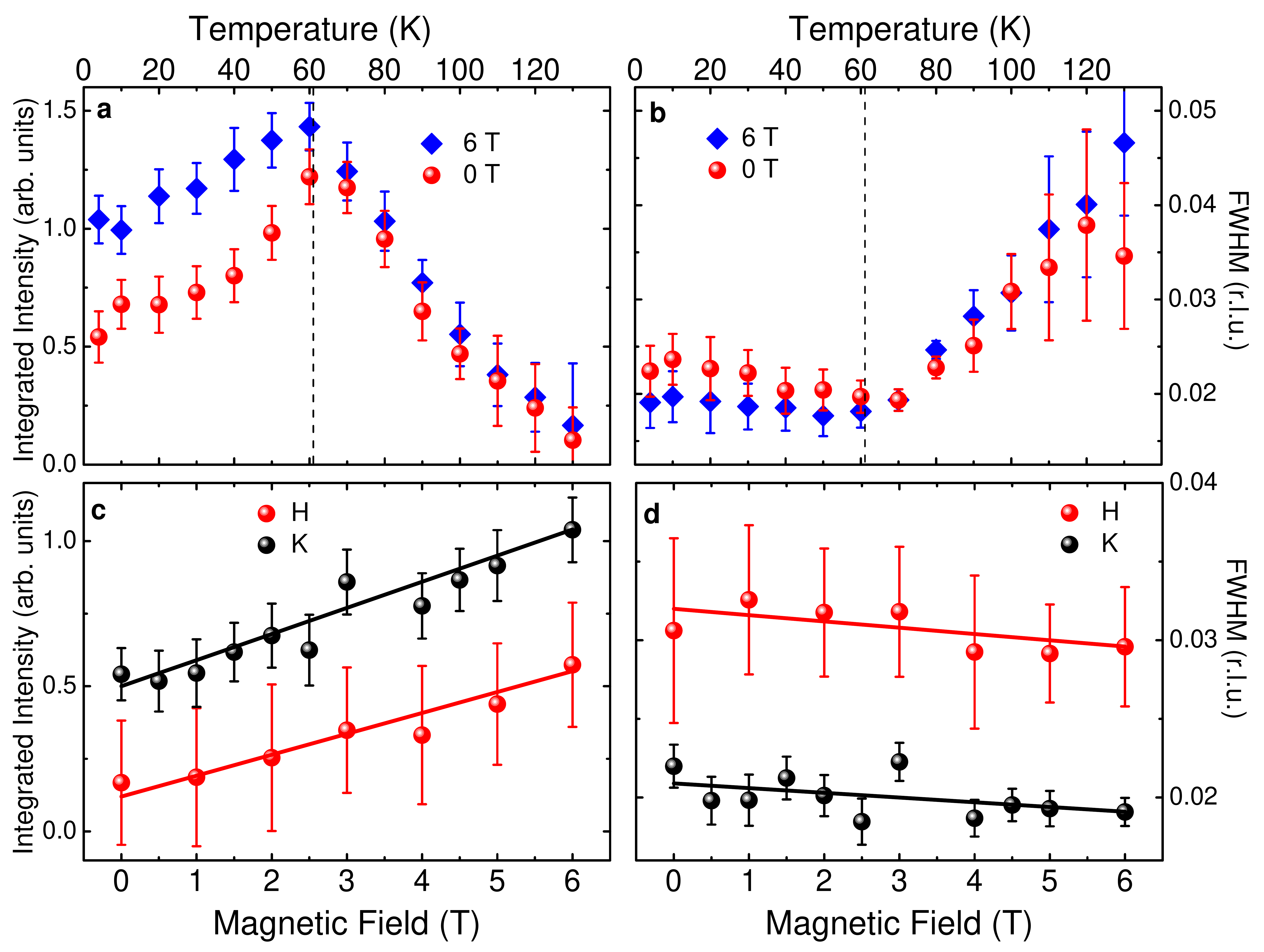}
\caption{(color online). (a) Temperature dependence of the integrated intensity of the CDW peak in ortho-II YBCO$_{6.55}$  along $(0, K, 0)$ for magnetic fields 0 and 6 T. (b) FWHM of the peak at two fields. The CDW is slightly more correlated below $T_c$ at high fields, but field independent above $T_c$ within the error bar. Magnetic field dependence of (c) the intensity and (d) the FWHM at 10 K. The integrated intensity depends linearly on magnetic field in both directions.
} \label{Fig2}
\end{figure}

Prior NMR~\cite{Wu_Nature2011} and quantum oscillation~\cite{Sebastian_RPP2012} experiments on samples similar to ours were performed in very high magnetic fields, sufficient to completely suppress superconducting long-range order. Even though it is not technically possible to-date to apply such high fields in RXS experiments, the behavior of the observed features in the lower fields that are already available is instructive (Fig. 2).
Below $T_c$, the integrated intensity of the peak increases linearly with field for both $H$ and $K$ directions, in agreement with previous hard x-ray work on ortho-VIII YBCO$_{6.67}$.~\cite{Chang_NaturePhysics2012} At 6 T, we observe an increase of the integrated intensity by a factor $\sim 2$ in both directions (either due to an enhanced CDW modulation amplitude, or due to an enhanced volume coverage of CDW domains), and we do not observe any sign of saturation. Although we cannot rule out a cross-over between the intensities of the modulations in the two directions and/or changes of $q_1$ or $q_2$ in higher fields leading to the commensurate pattern with charge modulation perpendicular to the chains suggested to account for NMR results,~\cite{Wu_Nature2011} no indication of such behavior has been observed up to 6 T. This is on the other hand consistent with recent sound velocity measurements that revealed field-induced static, biaxial order.~\cite{Leboeuf2012} Above $T_c$, the zero-field and high-field intensities match, in agreement with the notion that the field enhancement of the CDW fluctuations arises from a competition between CDW and superconducting instabilities. This is also supported by the weak decrease of the peaks' FWHM with field for $T < T_c$ (Fig~\ref{Fig2}-(d)).

It is interesting to compare the charge modulation in YBCO$_{6.55}$ revealed by our experiments to the one previously studied for YBCO$_{6.6}$, which exhibits a higher hole concentration. Both $q_1$ and $q_2$ are larger than the corresponding values in YBCO$_{6.6}$ (0.31 and 0.30, respectively). In the latter case, the difference is well outside the experimental error and indicates that the CDW modulation vector depends on the hole concentration in the CuO$_2$ layers, which is expected on general grounds but not observed outside the experimental error in our prior work.~\cite{Ghiringhelli_Science2012} Specifically, the shortening of the modulation vector with increasing doping level is qualitatively consistent with a scenario according to which the CDW wave vector reflects the nesting conditions of the Fermi surfaces arising from electronic states in the CuO$_2$ bilayers.~\cite{Andersen_PhysicaC1991, Eflimov_PRB2008} In this scenario, the strong in-plane anisotropy of the incommensurate peaks along $H$ and $K$, which contrasts with the more isotropic intensity distribution observed in YBCO$_{6.6}$, may reflect modified nesting conditions due to the influence of electronic states originating from the CuO chains and/or folding of the Fermi surfaces due to the long-range ortho-II superstructure. It is interesting to note that this anisotropy matches the one previously observed in the anomalous dispersion of copper-oxide phonon modes in YBCO$_{6+\delta}$.~\cite{Pintschovius_PRL2002,Raichle_PRL2011,Reznik_Nature2006}

Measurements of YBCO$_{6.55}$ and YBCO$_{6.6}$ crystals under the same experimental conditions further allowed us to compare the overall intensities of the incommensurate peaks in both crystals. The peak along $K$ is about $\sim 25$\% less intense in YBCO$_{6.55}$ than it is in YBCO$_{6.6}$; the intensity difference along $H$ is correspondingly larger. Apart from the modified Fermi surface geometry, the strong overall intensity reduction of the CDW peaks triggered by the small reduction of the hole content may reflect the influence of spin correlations, which are strongly enhanced in YBCO$_{6+\delta}$ with $\delta < 0.5$ where neutron scattering has revealed quasi-static incommensurate magnetic reflections.~\cite{Hinkov_Science2008,Haug_PRL2009,Haug_NJP2010} The CDW wave vector is unrelated to the wave vector characterizing magnetic order in samples of composition YBCO$_{6.45}$ (immediately adjacent in the phase diagram),~\cite{Hinkov_Science2008,Haug_NJP2010} suggesting that incommensurate spin and charge order are competing in the YBCO$_{6+\delta}$ system.

\begin{figure}
\includegraphics[width=0.95\linewidth]{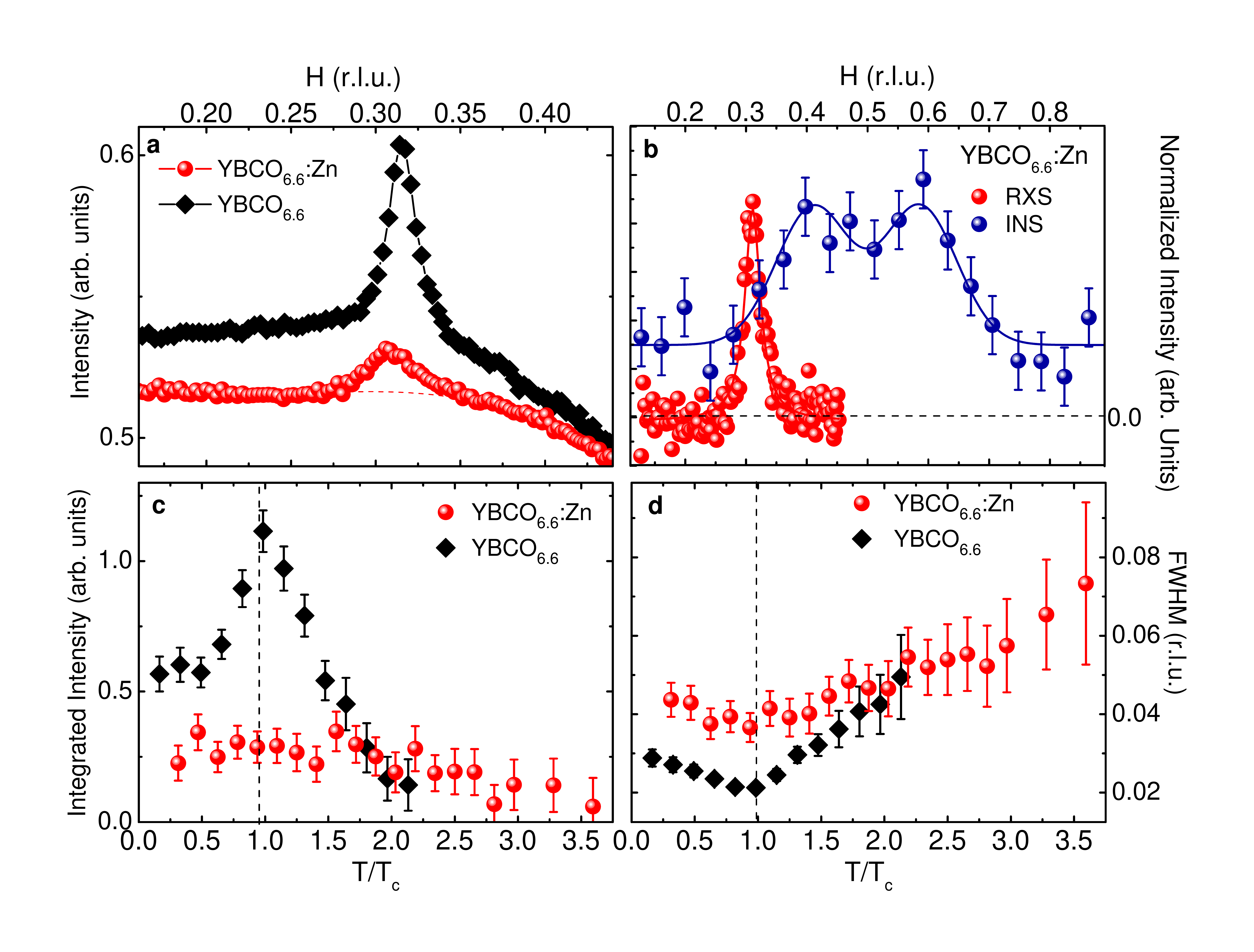}
\caption{(color online). (a) Comparison of the CDW peaks for pristine YBCO$_{6.6}$ and YBCO$_{6.6}$:Zn at their respective $T_c$. (b) Red dots: RXS data on YBCO$_{6.6}$:Zn from panel a, after subtraction of the background (dashed line in panel a). Blue dots: low-energy magnetic neutron scattering intensity obtained from inelastic neutron scattering  \cite{Suchaneck_PRL2010} at $T$=2 K on the mosaic of YBCO$_{6.6}$:Zn from which the crystal of this study was extracted. (c,d) Comparison of the integrated intensities and FWHM for pristine and Zn-substituted YBCO$_{6.6}$ as function of reduced temperature $T/T_c$ for each sample.
} \label{Fig3}
\end{figure}

To gain more insight into the interplay between spin and charge correlations, we have carried out RXS measurements on a YBCO$_{6.6}$ crystal in which 2\% of the in-plane Cu ions were replaced by spinless Zn impurities, which locally suppress superconductivity,~\cite{Bernhard_PRL96,Nachumi_PRL96} slow down the magnetic fluctuations,~\cite{Alloul_PRL91,Alloul_RMP2009} and induce an incommensurate magnetic structure with in-plane wave vector $(0.5 + \Delta, 0.5)$ akin to the one observed in pristine YBCO$_{6+\delta}$ at lower doping levels. The incommensurability parameter $\Delta \sim 0.1$ r.l.u. is consistent with the linear doping dependence observed in YBCO$_{6+\delta}$.~\cite{Suchaneck_PRL2010,Haug_NJP2010} We have reproduced the corresponding low-energy neutron scattering data in Fig. 3b.

The effect of spinless impurities on the CDW correlations is seen in Fig.~\ref{Fig3}(a), where we have plotted the measured intensity along $H$ for both pristine and Zn-substituted samples, close to their respective $T_c$. While Zn substitution does not modify the wave vector of the CDW peak significantly (consistent with the earlier conclusion that it does not modify the doping level,~\cite{Suchaneck_PRL2010}) its correlation length ($\xi \sim 20$~\AA~for $T=T_c$) and integrated intensity are drastically reduced in the YBCO$_{6.6}$:Zn sample. This is consistent with an inhomogeneous coexistence scenario in which the CDW correlations are suppressed in the area around the Zn impurities, where the incommensurate spin correlations are enhanced. Further support for this scenario comes from the absence of any straightforward relation between the wave vector of the incommensurate spin modulation measured by neutron scattering \cite{Suchaneck_PRL2010} and the charge modulation obtained here (Fig.~\ref{Fig3}(b)).

Figure~\ref{Fig3}(c,d) shows the temperature dependence of the integrated intensities and linewidths of the CDW peaks in the two samples. Whereas a maximum of the integrated intensity at $T_c = 32$ K is not clearly apparent in the Zn-substituted sample, the data do indicate a minimum of the linewidth, consistent with the  same competition between superconductivity and CDW correlations observed in Zn-free YBCO$_{6+\delta}$, albeit in a reduced volume fraction. The Zn-induced reduction of the CDW peak intensity (Fig. 3(a)) is in fact quantitatively consistent with the $\sim 70$\% reduction of the superfluid density reported earlier for a YBCO$_{6.6}$: 2\% Zn sample. \cite{Nachumi_PRL96}

In summary, our RXS data on YBCO$_{6.55}$ have revealed a nearly critical, biaxial CDW state with propagation vector similar to those of samples with higher $\delta$, ~\cite{Ghiringhelli_Science2012,Achkar_PRL2012,Chang_NaturePhysics2012} but with reduced amplitude and strong uniaxial anisotropy. They were taken on samples of composition identical to those that exhibit prominent quantum oscillations in high magnetic fields~\cite{Sebastian_RPP2012}, and the incisive information about the CDW order they have revealed now provides a solid basis for quantitative models of the Fermi surface reconstruction and electronic cyclotron orbits that underlie this phenomenon. The strong anisotropy of the CDW is in line with the previously observed trend of the correlated-electron system in strongly underdoped cuprates to form uniaxially modulated structures \cite{Hinkov_Science2008,Fujita_JPSJ2012,Kivelson_RMP2003, Votja_AdvPhys2009}. The CDW order we found is, however, inconsistent with the model proposed to explain recent high-field NMR data on samples of the same composition \cite{Wu_Nature2011}. This model should therefore be reconsidered.

X-ray and neutron scattering experiments have revealed three distinct low-temperature electronic phases in the YBCO$_{6+\delta}$ system. The corresponding order parameters magnetization density, charge density, and Cooper pairing amplitude prevail at low doping, at high doping and high magnetic field, and at high doping and low field, respectively. Rather than forming a coherent spin- and charge-modulated ``striped'' state, as in the 214 system,~\cite{Wilkins_PRB2011,Abbamonte_NatPhys2005,Fink_PRB2009,Fink_PRB2011,Hucker_PRB2011,Fujita_JPSJ2012} spin and charge order are strongly competing in YBCO$_{6+\delta}$. As a direct manifestation of this competition, we demonstrated that spinless Zn impurities substantially weaken CDW correlations in a YBCO$_{6.6}$ crystal, while at the same time nucleating incommensurate magnetic order. We further showed that an external magnetic field enhances the CDW correlations of YBCO$_{6.55}$ in the superconducting state, closely analogous to the field enhancement of the incommensurate spin correlations observed earlier in YBCO$_{6.45}$.~\cite{Haug_PRL2009} This implies that both spin and charge order compete against superconductivity, which is weakened by orbital depairing in the magnetic field. Taken together, these results thus indicate three competing electronic order parameters in YBCO$_{6+\delta}$.

\noindent \textit{Note added}. During the completion of this manuscript, we became aware of related results obtained on a ortho-II crystal by Blackburn {\it et al.}, using hard x-rays.~\cite{Blackburn}

\noindent \textit{Acknowledgement}. We acknowledge H. Alloul, M.-H. Julien, S. Kivelson, Y. Li, G. Sawatzky, and S. E. Sebastian for fruitful discussions.

%%%%%%%%%%%%%%%%%%%%%%BIBLIOGRAPHY%%%%%%%%%%%%%%%%%%%%%%%%%%%%%%%%%%%%%%%%%%%%%%%
%%%%%%%%%%%%%%%%%%%%%%%%%%%%%%%%%%%%%%%%%%%%%%%%%%%%%%%%%%%%%%%%%%%%%%%%%%%%%%%%%


\begin{thebibliography} {99}
\bibitem{Lee_RMP06} P. A. Lee, N. Nagaosa, and X. G. Wen, Rev. Mod. Phys. {\bf 78}, 17 (2006).
\bibitem{Hinkov_Science2008} V. Hinkov, D. Haug, B. Fauqu\'{e}, P. Bourges, Y. Sidis, A. Ivanov, C. Bernhard, C. T. Lin, and B. Keimer, Science {\bf 319}, 597 (2008).
\bibitem{Fujita_JPSJ2012} M. Fujita, H. Hiraka, M. Matsuda, M. Matsuura, J. M. Tranquada, S. Wakimoto, G. Y. Xu, and K. Yamada, J. Phys. Soc. Jpn. {\bf 81}, 011007 (2012).
\bibitem{Kivelson_RMP2003} S. A. Kivelson, I. P. Bindloss, E. Fradkin, V. Oganesyan, J. M. Tranquada, A. Kapitulnik, and C. Howald, Rev. Mod. Phys. {\bf 75}, 1201 (2003).
\bibitem{Votja_AdvPhys2009} M. Vojta, Adv. Phys. {\bf 58}, 699 (2009).
\bibitem{Parker_Nature2010} C. V. Parker, P. Aynajian, E. H. da Silva Neto, A. Pushp, S. Ono, J. Wen, Z. Xu, G. Gu, and A. Yazdani, Nature {\bf 468}, 677 (2010).
\bibitem{Wise_NPhys2008} W. D. Wise, M. C. Boyer, K. Chatterjee, T. Kondo, T. Takeuchi, H. Ikuta, Y. Wang, and E. W. Hudson, Nat. Phys. {\bf 4}, 696 (2008).
\bibitem{Koshaka_Science2007} Y. Kohsaka, C. Taylor, K. Fujita, A. Schmidt, C. Lupien, T. Hanaguri, M. Azuma, M. Takano, H. Eisaki, H. Takagi, S. Uchida, and J. C. Davis, Science {\bf 315}, 1380 (2007).
\bibitem{Wu_NatureCommunications2012} H. H. Wu, M. Buchholz, C. Trabant, C. F. Chang, A. C. Komarek, F. Heigl, M. v. Zimmermann, M. Cwik, F. Nakamura, M. Braden, and C. Sch\"{u}{\ss}ler-Langeheine, Nat. Commun. {\bf 3}, 1023 (2012).
\bibitem{Fink_PRB2009} J. Fink, E. Schierle, E. Weschke, J. Geck, D. Hawthorn, V. Soltwisch, H. Wadati, H.-H. Wu, H. A. D\"{u}rr, N. Wizent, B. B\"{u}chner, and G. A. Sawatzky, Phys. Rev. B {\bf 79}, 100502 (2009).
\bibitem{Fink_PRB2011} J. Fink, V. Soltwisch, J. Geck, E. Schierle, E. Weschke, and B. B\"{u}chner, Phys. Rev. B {\bf 83}, 092503 (2011).
\bibitem{Abbamonte_NatPhys2005} P. Abbamonte, A. Rusydi, S. Smadici, G. D. Gu, G. A. Sawatzky, and D. L. Feng, Nat. Phys. {\bf 1}, 155 (2005).
\bibitem{Wilkins_PRB2011} S. B. Wilkins, M. P. M. Dean, J. Fink, M. H\"{u}cker, J. Geck, V. Soltwisch, E. Schierle, E. Weschke, G. Gu, S. Uchida, N. Ichikawa, J. M. Tranquada, and J. P. Hill, Phys. Rev. B {\bf 84}, 195101 (2011).
\bibitem{Hucker_PRB2011} M. H\"{u}cker, M. v. Zimmermann, G. D. Gu, Z. J. Xu, J. S. Wen, G. Xu, H. J. Kang, A. Zheludev, and J. M. Tranquada, Phys. Rev. B {\bf 83}, 104506 (2011).
\bibitem{Kapitulnik_NJP2009} A. Kapitulnik, J. Xia, E. Schemm, and A. Palevski, New J. Phys. {\bf 11}, (2009).
\bibitem{Hosur_2012} Pavan Hosur, A. Kapitulnik, S.A. Kivelson, J. Orenstein, S. Raghu, arxiv 1212:2274.
\bibitem{Liang_PhysicaC2000} R. Liang, D. A. Bonn, and W. N. Hardy, Physica C {\bf 336}, 57 (2000).
%\bibitem{Karpinski_Nature1988} J. Karpinski, E. Kaldis, E. Jilek, S. Rusiecki, and B. Bucher, Nature {\bf 336}, 660 (1988).
%\bibitem{Zimmermann_PRB2003} M. v. Zimmermann, J. R. Schneider, T. Frello, N. H. Andersen, J. Madsen, M. K\"{a}ll, H. F. Poulsen, R. Liang, P. Dosanjh, and W. N. Hardy, Phys. Rev. B {\bf 68}, 104515 (2003).
\bibitem{Bangura_PRL2008} A. F. Bangura, J. D. Fletcher, A. Carrington, J. Levallois, M. Nardone, B. Vignolle, P. J. Heard, N. Doiron-Leyraud, D. LeBoeuf, L. Taillefer, S. Adachi, C. Proust, and N. E. Hussey, Phys. Rev. Lett. {\bf 100}, 047004 (2008).
\bibitem{Sebastian_RPP2012} S. E. Sebastian, N. Harrison, and G. G. Lonzarich, Rep. Progr. Phys. {\bf 75}, 102501 (2012).
\bibitem{Doiron_Nature2007} N. Doiron-Leyraud, C. Proust, D. LeBoeuf, J. Levallois, J.-B. Bonnemaison, R. Liang, D. A. Bonn, W. N. Hardy, and L. Taillefer, Nature {\bf 447}, 566 (2007).
\bibitem{Sebastian_PRL2012} S. E. Sebastian, N. Harrison, R. Liang, D. A. Bonn, W. N. Hardy, C. H. Mielke, and G. G. Lonzarich, Phys. Rev. Lett. {\bf 108}, 196403 (2012).
\bibitem{Laliberte_NatCom2011} F. Lalibert\'{e}, J. Chang, N. Doiron-Leyraud, E. Hassinger, R. Daou, M. Rondeau, B. J. Ramshaw, R. Liang, D. A. Bonn, W. N. Hardy, S. Pyon, T. Takayama, H. Takagi, I. Sheikin, L. Malone, C. Proust, K. Behnia, and L. Taillefer, Nat. Commun. {\bf 2}, 432 (2011).
\bibitem{Leboeuf_PRB2011} D. LeBoeuf, N. Doiron-Leyraud, B. Vignolle, M. Sutherland, B. J. Ramshaw, J. Levallois, R. Daou, F. Lalibert\'{e}, O. Cyr-Choini\`{e}re, J. Chang, Y. J. Jo, L. Balicas, R. Liang, D. A. Bonn, W. N. Hardy, C. Proust, and L. Taillefer, Phys. Rev. B {\bf 83}, 054506 (2011).
\bibitem{Timusk_RPP99} T. Timusk and B. W. Statt, Rep. Progr. Phys. {\bf 62}, 61 (1999).
\bibitem{Sebastian_PNAS2010} S. E. Sebastian, N. Harrison, M. M. Altarawneh, C. H. Mielke, R. Liang, D. A. Bonn, and G. G. Lonzarich, PNAS {\bf 107}, 6175 (2010).
\bibitem{Baek_PRB2012} S.-H. Baek, T. Loew, V. Hinkov, C. T. Lin, B. Keimer, B. B\"uchner, and H.-J. Grafe, Phys. Rev. B {\bf 86}, 220504(R) (2012).
\bibitem{Haug_NJP2010} D. Haug, V. Hinkov, Y. Sidis, P. Bourges, N. B. Christensen, A. Ivanov, T. Keller, C. T. Lin, and B. Keimer, New J. Phys. {\bf 12}, 105006 (2010).
\bibitem{Fong_PRB2000} H. F. Fong, P. Bourges, Y. Sidis, L. P. Regnault, J. Bossy, A. Ivanov, D. L. Milius, I. A. Aksay, B. Keimer, Phys. Rev. B {\bf 61}, 14773 (2000).
\bibitem{Ghiringhelli_Science2012} G. Ghiringhelli, M. Le Tacon, M. Minola, S. Blanco-Canosa, C. Mazzoli, N. B. Brookes, G. M. De Luca, A. Frano, D. G. Hawthorn, F. He, T. Loew, M. M. Sala, D. C. Peets, M. Salluzzo, E. Schierle, R. Sutarto, G. A. Sawatzky, E. Weschke, B. Keimer, and L. Braicovich, Science {\bf 337}, 821 (2012).
\bibitem{Achkar_PRL2012} A. J. Achkar, R. Sutarto, X. Mao, F. He, A. Frano, S. Blanco-Canosa, M. Le Tacon, G. Ghiringhelli, L. Braicovich, M. Minola, M. Moretti Sala, C. Mazzoli, R. Liang, D. A. Bonn, W. N. Hardy, B. Keimer, G. A. Sawatzky, and D. G. Hawthorn, Phys. Rev. Lett. {\bf 109}, 167001 (2012).
\bibitem{Chang_NaturePhysics2012} J. Chang, E. Blackburn, A. T. Holmes, N. B. Christensen, J. Larsen, J. Mesot, R. Liang, D. A. Bonn, W. N. Hardy, A. Watenphul, M. v. Zimmermann, E. M. Forgan, and S. M. Hayden, Nat. Phys. {\bf 8}, 871 (2012).
\bibitem{Wu_Nature2011} T. Wu, H. Mayaffre, S. Kr\"{a}mer, M. Horvatic, C. Berthier, W. N. Hardy, R. Liang, D. A. Bonn, and M.-H. Julien, Nature {\bf 477}, 191 (2011).
\bibitem{Haug_PRL2009} D. Haug, V. Hinkov, A. Suchaneck, D. S. Inosov, N. B. Christensen, C. Niedermayer, P. Bourges, Y. Sidis, J. T. Park, A. Ivanov, C. T. Lin, J. Mesot, and B. Keimer, Phys. Rev. Lett. {\bf 103}, 017001 (2009).
\bibitem{Alloul_RMP2009} H. Alloul, J. Bobroff, M. Gabay, and P. J. Hirschfeld, Rev. Mod. Phys. {\bf 81}, 45 (2009).
\bibitem{Suchaneck_PRL2010} A. Suchaneck, V. Hinkov, D. Haug, L. Schulz, C. Bernhard, A. Ivanov, K. Hradil, C.T. Lin, P. Bourges, B. Keimer, and Y. Sidis, Phys. Rev. Lett. {\bf 105}, 037207 (2010).
\bibitem{Lindemer_JACS1989} T.B. Lindemer, J. F. Hunley, J. E. Gates, A. L. Sutton, J. Brynestad,  C. R. Hubbard, and P. K. Gallagher, J. Amer. Ceram. Soc. {\bf 72}, 1775 (1989).
\bibitem{Liang_PRB2006} R. Liang, D. A. Bonn, and W. N. Hardy, Phys. Rev. B {\bf 73}, 180505 (2006).
\bibitem{Hawthorn_PRB2011} D. G. Hawthorn, K. M. Shen, J. Geck, D. C. Peets, H. Wadati, J. Okamoto, S. W. Huang, D. J. Huang, H. J. Lin, J. D. Denlinger, R. Liang, D. A. Bonn, W. N. Hardy, and G. A. Sawatzky, Phys. Rev. B {\bf 84}, 075125 (2011).
\bibitem{Leboeuf2012} D. LeBoeuf, S. Kr\"{a}mer, W. N. Hardy, R. Liang, D. A. Bonn, and C. Proust,Nat. Phys. {\bf 8}, 871 (2012)..
%\bibitem{magnet_xi} Note that the FWHM measured inside (Fig. \ref{Fig2}-b) and out of (Fig. \ref{Fig1}-d) the magnet chamber are slightly different. This is %caused by the fact that in the magnet, we used larger slits to compensate for the unfocussed beam, leading to a slight degradation of the momentum resolution. %For this reason we make no attempt to extract a correlation length from the measurement performed in the magnet, though we note that it increases slightly with %increasing field.
\bibitem{Eflimov_PRB2008} I. S. Elfimov, G. A. Sawatzky, and A. Damascelli, Phys. Rev. B {\bf 77}, 060504 (2008).
\bibitem{Andersen_PhysicaC1991} O. K. Andersen, A. I. Liechtenstein, O. Rodriguez, I. I. Mazin, O. Jepsen, A. P. Antropov, O. Gunnarsson, and S. Gopalan, Physica C {\bf 185-189}, 147 (1991).
\bibitem{Pintschovius_PRL2002} L. Pintschovius, W. Reichardt, M. Kl\"{a}ser, T. Wolf, and H. v. L\"{o}hneysen, Phys. Rev. Lett. {\bf 89}, 037001 (2002).
\bibitem{Reznik_Nature2006} D. Reznik, L. Pintschovius, M. Ito, S. Iikubo, M. Sato, H. Goka, M. Fujita, K. Yamada, G. D. Gu, and J. M. Tranquada, Nature {\bf 440}, 1170 (2006).
\bibitem{Raichle_PRL2011} M. Raichle, D. Reznik, D. Lamago, R. Heid, Y. Li, M. Bakr, C. Ulrich, V. Hinkov, K. Hradil, C. T. Lin, and B. Keimer, Phys. Rev. Lett. {\bf 107}, 177004 (2011).
\bibitem{Nachumi_PRL96} B. Nachumi, A. Keren, K. Kojima, M. Larkin, G. M. Luke, J. Merrin, O. Tchernyshov, Y. J. Uemura, N. Ichikawa, M. Goto, and S. Uchida, Phys. Rev. Lett. {\bf 77}, 5421 (1996).
\bibitem{Bernhard_PRL96} C. Bernhard, J. L. Tallon, C. Bucci, R. De Renzi, G. Guidi, G. V. M. Williams, and C. Niedermayer, Phys. Rev. Lett. {\bf 77}, 2304 (1996).
\bibitem{Alloul_PRL91} H. Alloul, P. Mendels, H. Casalta, J. F. Marucco, and J. Arabski, Phys. Rev. Lett. {\bf 67}, 3140 (1991).
\bibitem{Blackburn} E. Blackburn, J. Chang, M. H\"{u}cker, A. T. Holmes, N. B. Christensen, Ruixing Liang, D. A. Bonn, W. N. Hardy, M. v. Zimmermann, E. M. Forgan, and S. M. Hayden, Phys. Rev. Lett. {\bf 110}, 137004 (2013).

%\bibitem{Kimura_PRL2003} H. Kimura, M. Kofu, Y. Matsumoto, and K. Hirota, Phys. Rev. Lett. {\bf 91}, 067002 (2003).
%\bibitem{Huecker_2012} M. Huecker, M. v. Zimmermann, Z. J. Xu, J. S. Wen, G. D. Gu, J. M. Tranquada, arxiv:1212.3575.

%\bibitem{Scott_RMP1974} J. F. Scott, Reviews of Modern Physics {\bf 46}, 83 (1974).

\end{thebibliography}
\end{document}